\documentclass[12pt]{article}
\usepackage{cite,latexsym}

\newcommand{\bq}{{\bf q}}
\newcommand{\bmq}{{\bf |q|}}
\newcommand{\bk}{{\bf k}}
\newcommand{\bmk}{{\bf |k|}}
\newcommand{\bkq}{{\bf kq}}
\newcommand{\w}{\omega}
\newcommand{\W}{\Omega}

\newcommand{\ep}{\epsilon}

\begin{document}

\begin{titlepage}
\begin{flushright}
{September 1999}
\end{flushright}
\vskip 0.5 cm

\begin{center}
            {\large{\bf UNIQUENESS OF THE THERMAL EFFECTIVE
POTENTIAL }}

\vspace{0.6cm}

Georgios Metikas \footnote{E-mail address: g.metikas1@physics.oxford.ac.uk} 

\vspace{0.6cm} 

Department of Physics, Theoretical Physics,\\
University of Oxford, 1 Keble Road, Oxford OX1 3NP

\end{center}

\vspace{2cm}

\abstract{We discuss the use of derivative expansion techniques for
the construction of thermal effective potentials. We present a theory
for which the thermal bubble is analytic at the origin of the
momentum-frequency space, although the internal propagators in the
loop have the same mass. This means that, for this theory, the thermal
effective potential is uniquely defined. We then examine a slightly
different theory for which the thermal bubble displays the usual
non-analyticity at the origin and the thermal effective potential is
not uniquely defined. For this latter theory we compare our results 
 with those of other works in the literature which employ the
derivative expansion but find a uniquely defined thermal effective
potential. We raise several questions concerning the interchange of
the order of the perturbative and the derivative expansions, the
thermal generalization of some non-perturbative zero temperature
methods and the use of the periodicity of the external bosonic field.
 Finally, we re-examine the physical interpretation given to the
imaginary part of the thermal bubble in the literature.}

\end{titlepage}

\section{Introduction}

It is well-known that for most theories at finite temperature the
self-energy displays a non-analytic behaviour at the origin of the
momentum-frequency space \cite{WeldonMishaps,Das1}. This
non-analyticity manifests itself in a difference between the $ \left\{
q_0 \rightarrow 0, \bq \rightarrow 0 \right\} $ and  $ \left\{ \bq
\rightarrow 0, q_0 \rightarrow 0 \right\} $ limits of the self-energy,
where $q_0$ and $\bq $ are the components of the external momentum
$q_{\mu}=(q_0, \bq )$ and the component listed first goes to zero
first. The first limit leads to screening and the static potential
whereas the second limit has been used for the calculation of the
plasma frequency \cite{Kajantie,WeldonCovariant}. One may argue that
the two limits must differ since they refer to different physics
\cite{Kajantie}. However there is a problem with this argument. The
plasma frequency does not involve the $ \left\{ \bq \rightarrow 0, q_0
\rightarrow 0 \right\} $ limit but rather the $ q_0 \rightarrow \infty
$ limit \cite{Jackson}. It is because the latter limit is independent
of $\bq $ to lowest order that we appear to find the same result as with the former limit. This means that the physical significance of the $ \left\{ \bq \rightarrow 0, q_0 \rightarrow 0  \right\} $ limit is not well-understood yet.
 
The non-analyticity of the self-energy at the origin of the momentum-frequency space puts in jeopardy the construction of an effective potential based on the derivative expansion technique \cite{Zuk,Hott}. Historically, this problem was first pointed out in
the BCS theory context by Abrahams and Tsuneto \cite{Tsuneto}. Later it 
 was also seen to appear in a wide range of theories. In thermal QCD it occurs in the gluon \cite{Kalashnikov, WeldonCovariant} and in the quark self-energy \cite{Klimov, WeldonQuark}. Furthermore, it appears in all one-loop diagrams that have zero or two external quarks and any number of external gluons \cite{Braaten, FrenkelHigh}. The problem is also present  in the graviton self-energy \cite{RebhanAnalytical, RebhanCollective} and in higher-order graviton diagrams \cite{FrenkelHard}. Even in the much simpler case of interacting scalars the non-analyticity of the self-energy persists \cite{Fujimoto,WeldonMishaps,EvansZero}.  

The reason for this behaviour is that temperature effects give rise to Landau terms and these are responsible for the development of a new branch cut in the complex plane of the external momenta with a branch point at the origin, besides the usual one which is already present at zero temperature \cite{WeldonRules,Das1}. The usual branch cut exists for

\[ s= q_{0}^{2} - {\bmq }^{2} \geq  4m^{2} \]

\noindent and the new one for

\[ s= q_{0}^{2} - {\bmq}^{2}   \leq      0 . \]

An interesting remark is that, whenever the internal
propagators in a typical loop have different masses, the self-energy is analytic at the origin 
\cite{Vokos}. In this non-degenerate mass case the usual branch cut is
\[ s \geq (m_1 + m_2 )^2 \] 
and the new one is
\[ s \leq (m_1 - m_2 )^2 \] 
where $m_1$ and $m_2$ are the masses of the particles in the internal loop.
 The new branch point in not at the origin anymore and the problem disappears from this point, allowing thus the definition of a unique effective potential.
 However the non-degenerate mass case is of limited physical interest \cite{Vokos}.

We shall present a theory which exhibits a new and unexpected feature. The model of section 2 has self-energy which is analytic at the origin, although the mass is degenerate. 
In section 3 we show how subtle this new feature is and how the non-analyticity can develop, 
if we modify slightly our model by replacing the  parity 
conserving interaction term with a similar but parity 
violating one. Sections 2 and 3 are based to a great extend on a
previous work we did with M. Hott \cite{MetikasHott1}. Finally, in section 4, we compare our results with other in the literature.

\section{A new case} 

 We consider the following model
 \begin{equation}
 L[\bar{\psi}, \psi, \phi] = \bar{\psi} (i \!\not\! \partial - m) \psi -
 ig \bar{\psi} \gamma_5 \psi \phi + L_0[ \phi ] \label{L} 
 \end{equation}

 \noindent where $L_0[\phi]$ is the free Klein-Gordon Lagrangian. The boson is taken to be a pseudo-scalar quantity. 

We consider $\phi(x)$ to be an external field and
we want to obtain the one loop contribution to the effective action
which is given by
 
\begin{equation} 
\Gamma_{\mathit{eff}} [\phi]= 
 -i \ln{\frac{ { \mathrm{Det}} [i {S}^{-1}[\phi]]}{ { \mathrm{Det}} [i S^{-1}]}} 
\label{Det}
\end{equation}

 \noindent where $i S^{-1}[\phi]$ and $i {S}^{-1} $ are matrices whose
elements in coordinate representation are  

 \begin{eqnarray*}
 \langle x|iS^{-1}|y \rangle &=& (i \!\not\! \partial_x - m) \delta(x-y) \\
 \langle x|i {S}^{-1}[\phi]|y \rangle &=&(i \!\not\! \partial_x - m -
ig \gamma_5 \phi (x)) \delta(x-y). 
 \end{eqnarray*}

\noindent Since the external field depends on the coordinates, the
resulting functional determinants are not straightforward to
calculate. The matrices whose functional determinants we want to evaluate
are not diagonal in momentum or in coordinate space. However, progress can be made, if we rewrite (\ref{Det}) as: 

\begin{equation}
\Gamma_{\mathit{eff}}[\phi ] =
-i \mathrm{Tr} \ln{ \left[ 1- g \gamma_5 \phi (\hat{x})  S(\hat{p}) \right] }. \label{effaction1}
\end{equation}

\noindent Now we expand the above expression in powers of the coupling constant and show that the leading  contribution to the one-loop effective action is 

\begin{equation}
\Gamma^{(2)} = \frac{ig^2}{2} \int \frac{d^4q}{(2 \pi)^4}
\tilde{\phi}(-q) i \Pi (q) \tilde{\phi}(q)  \label{effaction2}
\end{equation}

\noindent where $ \tilde{\phi} (q) $ is the Fourier transformation of $
\phi (x) $ and 
\begin{equation}
i \Pi (q) = \int \frac{d^4k}{(2 \pi)^4} {\mathrm{tr}}  \left[ \gamma_5 \frac{1}{\not\! k 
 + \not\! q - m} \gamma_5 \frac{1}{ \not\! k -m} \right]. 
\end{equation}

\noindent  We note that $i \Pi(q) $ is just the self-energy
 bubble diagram for the boson which, after performing the trace, is given by  
\begin{equation}
 i \Pi(q)= - 4 \int \frac{d^4k}{ (2 \pi)^4 }
 \frac{ k^2 + k^{\mu}q_{\mu} - m^2}{[(k+q)^2 - m^2][ k^2 - m^2]} . \label{new}
 \end{equation}
\noindent This is one typical diagram that usually has a
non-analytic behaviour in the limit of vanishing external momenta
but we are going to show that this is not the case here. We keep
this intermediate expression, because it will help us to show in the
next section how the non-analyticity can develop in the scalar-coupling model.

%It is worth mentioning that the leading contribution to the one-loop
%effective action can also be written as
% 
%\[ \Gamma^{(2)} = \frac{ig^2}{2} \int {d^4 x}
% \; \phi (x) i \Pi (q) \phi (x) , \]
% 
%\noindent where $q_\mu$ is to be understood as a derivative operator
%acting on the field to the right. This result can be obtained by
%inserting complete sets of eigenvectors of momentum and position
%operators. The external field is an operator acting on coordinate
%states which at finite temperature are thermal states. Then the
%eigenfunction $\phi(x)$ has to be interpreted as being thermalized
%itself. This seems to us to be the reason why the approach given in
%\cite{EvansDerivative,EvansUnique} is not in agreement with what is
%usually expected for the effective action at finite temperature.
 
Applying the usual finite temperature techniques to (\ref{new}), we find the following expression for the thermal bubble diagram.

 \begin{eqnarray}
 &&  \Pi (q_{0}, \bq ) = \int \frac{d^3 \bk }{ (2 \pi )^3 } \
 \frac{1}{ 2 \w \W } \left\{ 2 \w \left[ \tanh{ \frac{  \beta ( \W +
q_0 ) }{2}} + \tanh{ \frac{ \beta ( \W - q_0 )}{2}} \right]
\right. \nonumber \\
 &&+ \frac{1}{ \W + \w - q_0 } \left[ [ \w q_0 + \bkq ] \tanh{ \frac{
 \beta \w }{2} } + [ q_0^2 - \W q_0 + \bkq ] \tanh{ \frac{\beta ( \W - q_0)}{2}} \right]   \nonumber \\ 
 &&+ \frac{1}{ \W + \w + q_0 }
  \left[ [ - \w q_0 + \bkq ] \tanh{ \frac{ \beta \w }{2}} + [q_0^2 + \W q_0
 + \bkq ] \tanh{ \frac{ \beta ( \W + q_0) }{2}} \right] \nonumber \\
 &&+ \frac{1}{ \W - \w + q_0 }
 \left[ [ \w q_0 + \bkq ] \tanh{ \frac{ \beta \w }{2}} - [ q_0^2 + \W q_0
 + \bkq ] \tanh{ \frac{ \beta ( \W + q_0 )}{2}} \right] \nonumber \\
 && \left. + \frac{1}{ \W - \w - q_0 }
 \left[ [- \w q_0 + \bkq ] \tanh{ \frac{ \beta \w }{2}} - [ q_0^2 - \W q_0
 + \bkq ] \tanh{ \frac{ \beta ( \W - q_0)}{2}} \right] \right\} \nonumber \\ 
\label{thermal bubble before} 
\end{eqnarray}

\noindent where 
\begin{equation}
\w = \sqrt{ \bk ^2 + m^2} \hspace{2cm} \W = \sqrt{
( \bk + \bq )^2 + m^2 } \hspace{2cm} q_0= i \frac{2 \pi n}{\beta}, \; n= \mathrm{integer}.
\end{equation}
\noindent From the above definition of $q_0$ follows that
\begin{equation}
e^{ \beta q_0 } = 1 \label{crucial}
\end{equation}
\noindent and consequently $q_0$ disappears from all the hyperbolic
 tangents of (\ref{thermal bubble before}).

\begin{eqnarray}
 &&  \Pi (q_{0}, \bq ) =  \int \frac{d^3 \bk }{ (2 \pi )^3 } \
 \frac{1}{ 2 \w \W } \left\{ 4 \w \tanh{ \frac{  \beta \W }{2}} +
\right. \nonumber \\
 &&+ \frac{1}{ \W + \w - q_0 } \left[ [ \w q_0 + \bkq ] \tanh{ \frac{
 \beta \w }{2} } + [ q_0^2 - \W q_0 + \bkq ] \tanh{ \frac{\beta \W }{2}} \right]   \nonumber \\ 
 &&+ \frac{1}{ \W + \w + q_0 }
  \left[ [ - \w q_0 + \bkq ] \tanh{ \frac{ \beta \w }{2}} + [q_0^2 + \W q_0
 + \bkq ] \tanh{ \frac{ \beta \W }{2}} \right] \nonumber \\
 &&+ \frac{1}{ \W - \w + q_0 }
 \left[ [ \w q_0 + \bkq ] \tanh{ \frac{ \beta \w }{2}} - [ q_0^2 + \W q_0
 + \bkq ] \tanh{ \frac{ \beta \W }{2}} \right] \nonumber \\
 && \left. + \frac{1}{ \W - \w - q_0 }
 \left[ [- \w q_0 + \bkq ] \tanh{ \frac{ \beta \w }{2}} - [ q_0^2 - \W q_0
 + \bkq ] \tanh{ \frac{ \beta \W }{2}} \right] \right\}. \nonumber \\  
\label{thermal bubble}
 \end{eqnarray}

%\noindent where 
%\begin{equation}
% \w = \sqrt{ \bk ^2 + m^2} \hspace{2cm} \W = \sqrt{
%( \bk + \bq )^2 + m^2 } \hspace{2cm} q_0= i \frac{2 \pi m}{\beta}
%\end{equation}
%\noindent We note that for the derivation of (\ref{thermal bubble}) we have used the identity 
%\begin{equation}
%e^{ \beta q_0 } = 1 \label{crucial}
%\end{equation}
%\noindent which follows from $q_0=i 2 \pi m/ \beta $.  

Even before performing the angular integration, we can have a first
naive indication that the zero-momentum limit of (\ref{thermal bubble}) does not display the usual
non-uniqueness problem. The two successive limits are 

\begin{eqnarray*}
\Pi (0, \bq ) & = &  \int \frac{d^3 \bk }{ (2 \pi )^3 } \; \frac{1}{
\w \W } 
\left\{ 2 \w \tanh{\frac{  \beta \W }{2} }  
+  \frac{ \bkq }{ \W + \w } \left[ \tanh{\frac{ \beta \w }{2} } +
\tanh{\frac{ \beta \W }{2} } \right] \right.  \nonumber \\ 
& + &  \left.  \frac{\bkq}{\W - \w }
\left[ \tanh{ \frac{ \beta \w }{2}} - \tanh{\frac{ \beta \W }{2}} 
\right] \right\} , 
\end{eqnarray*}

\noindent which can be checked to give
 
\begin{equation}
 \lim_{ {\bmq} \rightarrow 0} \Pi (0, {\bf q}) =  \frac{1}{{\pi}^2}
\int_{|m|}^{ \infty }
 d \w \sqrt{{\w}^{2}-m^{2}}\tanh{ \frac{ \beta \w }{2}} \; .
\end{equation}

\noindent Similarly we find

\begin{eqnarray*}
  \Pi (q_{0}, 0) & = & \int \frac{d^3 \bk }{ (2 \pi )^3 } \
 \frac{1}{ 2 \w \W } \left\{ 4 \w \tanh{ \frac{  \beta \W }{2}} +
\right. \nonumber \\
 & + & \left. {q_{0}}^{2} \left[ \frac{1}{ 2 \w - q_0 } + \frac{1}{ 2 \w +
q_0} \right] \tanh{ \frac{ \beta \w }{2}}  \right\}  \nonumber \\
& \stackrel{q_{0} \rightarrow 0}{ \longrightarrow } & \frac{1}{{\pi}^2}
\int_{|m|}^{ \infty } d \w \sqrt{{\w}^{2}-m^{2}} \tanh{ \frac{\beta
\w}{2}}.
\end{eqnarray*}

\noindent We conclude that the limits coincide. Moreover, the only term
that contributes to the unique result is the first one in the
integrand of equation (\ref{thermal bubble}) and those proportional to
Landau terms - the last two terms inside the integrand - vanish in this limit. 

A more general way of seeing that the limits are the same is to
perform the angular integration and then use the 
parameterization $ q_0 = a \bmq $, where $a$ can be
 any real number, and find the limit of $ \Pi (a \bmq , \bmq ) $ as $
\bmq \rightarrow 0 $. If the limit is independent of $a$, we have a
strong indication that the function
is analytic at the origin, {\it i.e.} it does not depend on the way one
approaches the origin \cite{WeldonMishaps, Vokos}. Before doing so we recast
equation (\ref{thermal bubble}) in a more convenient form by means of
the transformation $ \bk \rightarrow  - (\bk + \bq) $ wherever the
integrand contains $\tanh{\frac{\beta \W}{2}} $. Then we find  

\begin{eqnarray}
 \Pi(q_0, \bq ) & = & \int \frac{d^3 \bk }{ (2 \pi )^3 } \left\{ \frac{2}{ \w }
  \tanh{ \frac{\beta \w }{2} } + ( q_0^2 - \bq ^2 ) \tanh{ \frac{ \beta \w
 }{2} } \right. \nonumber \\
 & \times & \left.\frac{1}{2 \w \W} \left[ \frac{1}{ q_0 + \W + \w } - \frac{1}{
 q_0 - \W - \w } + \frac{1}{ q_0 + \W - \w } - \frac{1}{ q_0 - \W + \w
 } \right] \right\}. \label{selfenergy}
 \end{eqnarray}
 \noindent One can note that at $T=0$ the Landau terms cancel each other, as expected.
 We change variables from $ \cos{\theta} $ to $ \W $ and perform the
 integration over $ \W $. The result is 

\begin{eqnarray}
\Pi ( q_0 , \bmq ) & = &  \frac{1}{{\pi}^2}
\int_{|m|}^{ \infty } d \w \sqrt{{\w}^{2}-m^{2}} \tanh{ \frac{ \beta\w}{2} }   \nonumber \\
& + &  \frac{ q_0^2 - \bmq ^2 }{2 \bmq } \int_{|m|}^{ \infty } \frac{
d \w }{ (2\pi)^2 } \tanh{ \frac{\beta \w}{2} } [ L1 + L2 + L3+ L4 ]
\label{angular thermal bubble}
\end{eqnarray}
 
\noindent where 
\begin{eqnarray*}
 L1( q_0, \bmq ) = \ln{ \frac{ \W _{+} + \w + q_0 }{ \W_{-} + \w +  q_0}} &&
 L2( q_0, \bmq ) = \ln{ \frac{ \W _{+} + \w - q_0 }{ \W_{-} + \w -  q_0}} \\
 L3( q_0, \bmq ) = \ln{ \frac{ \W _{+} - \w + q_0 }{ \W_{-} - \w +  q_0}} &&
 L4( q_0, \bmq ) = \ln{ \frac{ \W _{+} - \w - q_0 }{ \W_{-} - \w -  q_0}} 
\end{eqnarray*}
 
\noindent with \[ \W _{+}= \sqrt{ (\bmk + \bmq )^2 + m^2 }
 \hspace{2cm}  \W _{-}= \sqrt{ (\bmk - \bmq )^2 + m^2 }. \]
\noindent If $q_0$ is
made complex and continuous, the only poles or zeros of the sum of L's
in (\ref{angular thermal bubble})
occur for $q_0$ on the real axis. It is perfectly appropriate to have
singularities on the real axis. Thus the analytic extension of
(\ref{angular thermal bubble}) is trivially obtained by letting $q_0$ be real. 
There are three self-energies on the real axis.
\begin{eqnarray*}
\Pi_R ( q_0, \bmq ) = \Pi(q_0 + i \ep, \bmq ) & \Pi_A(q_0, \bmq ) = \Pi(q_0 - i \ep, \bmq ) & \Pi_F(q_0, \bmq ) = \Pi(q_0 + i \ep q_0 , \bmq )
\end{eqnarray*}
\noindent where $\ep \rightarrow 0^+ $. The real parts of these self-energies coincide whereas the imaginary parts are related according to
\begin{equation}
\mathrm{Im}\Pi_R=-\mathrm{Im}\Pi_A = \tanh{(\frac{\beta q_0}{2})} \mathrm{Im}\Pi_F.
\end{equation}
\noindent Following \cite{WeldonMishaps} we shall not concern ourselves with the Feynman self-energy. Using the fact that $\ep$ is infinitesimal, the real part of the self-energy can be shown to be 
\begin{eqnarray}
\mathrm{Re}\Pi ( q_0 , \bmq ) & = &  \frac{1}{{\pi}^2}
\int_{|m|}^{ \infty } d \w \sqrt{{\w}^{2}-m^{2}} \tanh{ \frac{ \beta\w}{2} }   \nonumber \\
& + &  \frac{ q_0^2 - \bmq ^2 }{2 \bmq } \int_{|m|}^{ \infty } \frac{ d \w }{ (2\pi)^2 } \tanh{ \frac{\beta \w}{2} } [ \mathrm{Re}L1 + \mathrm{Re}L2 + \mathrm{Re}L3 + \mathrm{Re}L4 ]
\end{eqnarray}

\noindent where 

\begin{eqnarray*}
 \mathrm{Re}L1( q_0, \bmq ) = \ln{\left| \frac{ \W _{+} + \w + q_0 }{ \W_{-} + \w +  q_0} \right|} &&
 \mathrm{Re}L2( q_0, \bmq ) = \ln{\left| \frac{ \W _{+} + \w - q_0 }{ \W_{-} + \w -  q_0} \right| } \\
 \mathrm{Re}L3( q_0, \bmq ) = \ln{\left| \frac{ \W _{+} - \w + q_0 }{ \W_{-} - \w +  q_0} \right|} &&
 \mathrm{Re}L4( q_0, \bmq ) = \ln{\left| \frac{ \W _{+} - \w - q_0 }{ \W_{-} - \w -  q_0} \right| }. 
\end{eqnarray*} 

\noindent We note that the real part of the self-energy is even under
$q_0 \rightarrow -q_0$, since it can be written as a function of
$q_0^2$, if we combine the logarithms. It is also even under ${\bf q}
\rightarrow {\bf -q}$, since it depends only on $\bmq $.
\\
\noindent Now we turn to the imaginary part which we calculate
according to \cite{WeldonRules}, making use of the following form of
the delta function

\begin{equation}
\delta(x) = \frac{i}{2 \pi} \lim_{ \ep \rightarrow 0^+} \left[ \frac{1}{x+i \ep} -
\frac{1}{x -i \ep} \right].
\end{equation}

\noindent The imaginary part is

\begin{eqnarray}
&& \mathrm{Im} \Pi_R = - \mathrm{Im} \Pi_A =  \frac{1}{2} \int \frac{d^3 {\bf
k}}{(2 \pi )^2 } \ \frac{1}{2 \w \W} \ \tanh{ (\frac{ \beta \w }{2} ) }
 \ (q_0^2- \bq ^2) \nonumber \\ && \left[ \delta(q_0 + \W + \w ) -
\delta(q_0 -\W -\w ) + \delta(q_0 + \W - \w ) - \delta( q_0 - \W + \w ) \right].
\end{eqnarray}

\noindent We note that it is odd under $ q_0 \rightarrow - q_0 $.
 However it is even under $ {\bf q} \rightarrow - {\bf q} $, because
we can simultaneously change the integration variable $ {\bf k} \rightarrow
- {\bf k} $ . This means that, unlike the real part, the imaginary
part of the retarded or advanced thermal self-energy 
does not contribute to the effective action. As we can see
from (\ref{effaction2}), the integrand of the effective action is $
\tilde{\phi}(q_0, \bq ) \ \tilde{\phi}(- q_0, - \bq ) \ \Pi(q_0, \bq ) $
and therefore the contribution  $ \tilde{\phi}(q_0, \bq ) \
\tilde{\phi}( - q_0, - \bq ) \ \mathrm{Im}\Pi(q_0, \bq ) $ is odd under
the combined transformations $ q_0 \rightarrow -q_0 $ and $ \bq
\rightarrow - \bq $ and vanishes, when integrated over $d^4q$.

\noindent  We proceed to the parameterization $ q_0 = a \bmq
$ and examine the behaviour of the real part of the self-energy as $
\bmq \rightarrow 0$. The limits of two of the regular terms
$\mathrm{Re}L1$ and $\mathrm{Re}L2$ are independent of $a$, as they should be. We can see that

 \[ \lim_{ \bmq \rightarrow 0} (a^2 -1)\bmq \mathrm{Re}L1 = 0
\hspace{2cm} \lim_{ \bmq \rightarrow 0} (a^2 -1)\bmq \mathrm{Re}L2  = 0 . \]

\noindent  What is quite unexpected is that, for this particular
model, the contributions coming from the Landau terms, $\mathrm{Re}L3$
and $\mathrm{Re}L4$, vanish independently of $a$, that is

 \[ \lim_{
 \bmq \rightarrow 0 } (a^2 -1)\bmq \mathrm{Re}L3 = 0   
  \hspace{2cm} \lim_{ \bmq \rightarrow 0} (a^2 -1)\bmq \mathrm{Re}L4 = 0.\]

\noindent In other words, although the Landau terms are not
well-behaved at the origin of momentum space, a unique effective
potential up to second order in the coupling constant can be defined
here thanks to the kinetic term in the numerator of equation
(\ref{selfenergy}), namely  $q_0^2 - {\bf q}^2$. This is an
interesting result but this kinetic term does not always appear in
bubble diagrams as we are going to see in the next section.  
In the present case the one-loop, $g^2$ order contribution to the
effective potential is
 
 \begin{eqnarray}
 V_{\mathit{eff}}^{(2)} &=& - \frac{i g^2}{2} i \mathrm{Re}\Pi(0,0)  \phi^{2} \nonumber \\
 \mathrm{Re}\Pi(0,0) &=& \frac{1}{ \pi ^2} \int_{|m|}^{ \infty } d \w \sqrt{ \w ^2
 - m^2 } \tanh{ \frac{ \beta \w }{2} }.  
 \end{eqnarray}
 
\noindent This result for the effective potential coincides with the
one we would have found, had we not added to the real $ q_0 $ the
infinitesimal imaginary part which corresponds to physical boundary
conditions \cite{MetikasHott1}. 
The next order in the derivative expansion is non-analytic since the
derivatives of the Landau terms become dominant and the derivative expansion breaks down.
 
\section{ A usual case }

In this section we shall consider a model whose only difference from
the one which we examined in the previous section is that its
interaction term does not contain the $\gamma_{5}$ matrix. 

\begin{equation}
 L^{\prime}[\bar{\psi}, \psi, \phi] = \bar{\psi} (i \!\not\! \partial
- m) \psi - ig \bar{\psi} \psi \phi + L_0[ \phi ]. \label{Lusual} 
 \end{equation}

\noindent As we shall soon see, this simple modification of the
interaction term has far-reaching consequences as far as the analytic properties of the
thermal self-energy are concerned. Starting from (\ref{Lusual}) and
following the procedure of the previous section we find that the
one-loop effective action is

\begin{equation}
\Gamma^{\prime}_{\mathit{eff}}[\phi ] =
-i \mathrm{Tr} \ln{ \left[ 1- g \phi (\hat{x})  S(\hat{p}) \right] }. 
\label{effaction1usual}
\end{equation}

\noindent and the self-energy bubble is given by

\begin{equation}
 i \Pi^{'}(q)=  4 \int \frac{d^4k}{ (2 \pi)^4 }
 \frac{k^2 + k^{\mu}q_{\mu} + m^2}{[(k+q)^2 - m^2][ k^2 - m^2]} 
\label{usual}
\end{equation}
 
\noindent which can be written as

\[ i \Pi^{'}(q) = - i \Pi(q) + i \Pi^{''}(q), \]

\noindent where

\begin{eqnarray*}
 i \Pi^{''}(q) = 4 \int \frac{d^4k}{(2 \pi)^4} \frac{2 m^2}{[(k+q)^2 -
m^2][ k^2 - m^2]}.
\end{eqnarray*}	     

As we saw in the previous section $ \mathrm{Re} \Pi(a\bmq, \bmq )$ does not depend
on $a$, when $\bmq \rightarrow 0$. We will see that $\mathrm{Re}
\Pi^{''}(a\bmq ,\bmq ) $ does. We have

\begin{eqnarray}
\Pi^{''}(q_0, \bq )= - m^2 \ \int \frac{ d^3 \bk }{(2 \pi)^3} \
\frac{1}{\w \W } && \ \left\{ \left[ \frac{1}{\W +\w - q_0} + \frac{1}{\W
-\w +q_0}  \right] \ \tanh{\frac{\beta \w}{2}} \right. \nonumber \\
&& + \left[  \frac{1}{\W + \w + q_0} + \frac{1}{\W - \w - q_0} \right] \
 \tanh{ \frac{\beta \w }{2}}  \nonumber \\
&& + \left[ \frac{1}{\W +\w + q_0} - \frac{1}{\W
-\w  + q_0} \right] \ \tanh{\frac{\beta (\W + q_0)}{2}} \nonumber \\
&& \left. + \left[ \frac{1}{\W +\w - q_0} - \frac{1}{\W
 - \w  - q_0} \right] \ \tanh{\frac{\beta (\W - q_0)}{2}} \right\}
\label{usual thermal bubble 1}
\end{eqnarray}

\noindent which after using (\ref{crucial}) and applying the
transformation $ \bk \rightarrow -( \bk + \bq ) $ becomes

\begin{eqnarray}
&& \Pi^{''}(q_0,\bq )= -2 m^2 \ \int \frac{d^3 \bk }{(2 \pi )^3 } \
 \frac{1}{ \w \W } \ \tanh{ \frac{\beta \w }{2} } \nonumber \\
& \times & \left\{ \frac{1}{q_0 + \W + \w } - \frac{1}{q_0 - \W -\w } +
 \frac{1}{q_0 + \W - \w } - \frac{1}{q_0 - \W + \w }  \right\}.
\label{usual thermal bubble 2}
\end{eqnarray}

\noindent Performing the angular integration, setting $ q_0 =a \bmq $ and following the steps of the previous paragraph yields the effective potential 

\begin{eqnarray}
&& (V_{\mathit{eff}}^{(2)})^{''} = \frac{g^2}{2} \mathrm{Re} \Pi^{''} (0,0) \ \phi^2 \nonumber \\
&& \mathrm{Re} \Pi^{''} (0,0) = \frac{1}{ \pi ^2 } \ \int_{|m|}^{ \infty } d \w \ 
 \sqrt{\w^2 - m^2} \ \tanh{ \frac{\beta \w }{2}} \ \left\{
 \frac{m^2}{ \w ^2} + \frac{m^4}{(w^2 - m^2) \w ^2 - a^2 \w ^4 } \right\}.
\end{eqnarray}

% \begin{eqnarray}
%&& \lim_{\bmq \rightarrow 0} \Pi^{''}(a \bmq ,\bmq ) =  \frac{m^2}{\pi^2}
%\int_{|m|}^{ \infty } d \w \left\{ \frac{ \sqrt{ \w^2 - m^2 } }{\w^2} \tanh{
% \frac{ \beta \w}{2} } - \right. \nonumber \\
%&& -  \left. \frac{\beta}{2 \w}    \cosh^{-2}{ \frac{\beta \w}{2} }
%\left[ \sqrt{\w^2 - m^2} - \frac{\w a}{2} \ln{ \frac{| \w a + \sqrt{
%\w^2- m^2 }| }{|\w a - \sqrt{ \w^2- m^2 }| } } \right] \right\} .  
% \end{eqnarray}	     

\noindent As in the previous section, the imaginary part does not
contribute to the effective action.
Therefore the total effective potential for the theory of
this section is

\begin{eqnarray}
&& (V_{\mathit{eff}}^{(2)})^{'} =  \frac{g^2}{2} \phi^2 \left[ - \Pi(0,0) +
\Pi^{''}(0,0) \right] \nonumber \\
&& =  - \frac{g^2}{ 2 \pi^2 } \ \int_{|m|}^{\infty}
dw \tanh{ \frac{\beta \w }{2} } \ \sqrt{\w ^2 -m^2} \ \left\{ 1-
\frac{m^2}{\w ^2} - 
 \frac{m^4}{(\w ^2 -m^2) \w ^2 - a^2 \w ^4}  \right\} \ \phi ^2.  
\label{usual effective potential}
\end{eqnarray}

\noindent This result for the effective potential coincides with the
one we would have found, had we not added to the real $q_0$ the
infinitesimal imaginary part which corresponds to physical boundary
conditions \cite{MetikasHott1}.

This effective potential is not uniquely defined, because it depends
on $a$ which can take any real value. Comparing (\ref{new}) to
(\ref{usual}), we see that dropping
$\gamma_5$ from the interaction resulted in changing the relative sign
between the momentum terms and $m^2$ in the numerator. This slight
change was enough to allow for the development of a self-energy which
is non-analytic at the origin. 
 
\section{Comparison with other works}

The purpose of this section is to compare our results with others in the
literature.

\subsection{Comparison with Dolan and Jackiw}

For the theory examined in section (3) the one-loop effective potential at order $g^2$ is
given by (\ref{usual effective potential}).
%\begin{eqnarray}
%&& V^{(2)}_{\mathit{eff}} =  - \frac{g^2}{2 \pi^2} \int_{|m|}^{\infty} d\w \left\{
%\sqrt{\w^2 - m^2} \left( 1 - \frac{m^2}{\w^2} \right)
%\tanh{ \frac{ \beta \w }{2} } \; + \right. \nonumber \\
%&& \left. \frac{\beta }{2 \w } \cosh^{-2}{\frac{\beta \w }{2} } \left[
%\sqrt{\w^2 - m^2} - \frac{\w a}{2} \ln{ \frac{| \w a + \sqrt{ \w^2-
%m^2}| }{| \w a - \sqrt{\w^2- m^2} | } } \right] \right\} \;
%\phi^{2}(x).  \label{our}
%\end{eqnarray}
\noindent In \cite{Dolan} the same theory was considered and, setting
the external field to be constant, Dolan and Jackiw obtained the
following exact expression for the one-loop effective potential

\begin{equation}
V_{\mathit{eff}}= - \frac{2}{\pi^2} \int_{|m|}^{\infty} d\w \; \w \; \sqrt{\w^2
-m^2} \; \left[ \frac{E}{2} + \frac{1}{\beta} \ln{(1 + e^{ - \beta E})}
\right], \label{exact effective potential}
\end{equation}

\noindent where 

\[ E= \left[ \w^2 - m^2 + (m + g \phi)^2 \right]^{1/2}. \] 

\noindent We are interested in the contribution at the second order in the
coupling constant which is

\begin{eqnarray}
V^{(2)}_{\mathit{eff}} & = & - \frac{g^2}{2 \pi^2} \int_{|m|}^{\infty} d\w
 \; \sqrt{\w^2 - m^2} \; \left\{ \left( 1 - \frac{m^2}{\w^2} \right)
\tanh{\frac{\beta \w}{2} }+ \right. \nonumber \\
& + & \left. \frac{ m^2 \beta}{2 \w}  \cosh^{-2}{ \frac{\beta \w }{2}
} \right\} \; \phi^{2}. \label{dolan}
\end{eqnarray}

\noindent 
%If we set $a=0$ in equation (\ref{our}), it reduces to expression
%(\ref{dolan}), which means that the result derived by Dolan and Jackiw
%is valid only in one of the infinite number of ways of approaching the
%origin, namely in the case where we take $q_0 \rightarrow 0 $ first
%and then $ \bq \rightarrow 0$. 
One can  reproduce equation (\ref{dolan}) by setting $(q_0, \bq ) = (0, 0) $ in
formula (\ref{usual}) and then performing the Matsubara sum. However, 
the correct thing to do is to perform the sum first so that the explicit
form of the self-energy as a function of $q_0$ and $\bmq $ is obtained. Then the
behaviour of this function can be investigated in the limit $(q_0, \bq ) \rightarrow  (0, 0) $. We
therefore conclude that the non-perturbative method employed in
\cite{Dolan} is not generally equivalent to the perturbative calculation, because it fails
to take into account the non-analyticity which appears at the origin
of the space of external momenta.  

It would be interesting to investigate further why our result doesn't
coincide with the one derived in \cite{Dolan}. Instead of just
comparing the final results we shall try to find out in which stage of
the calculation the difference between \cite{Dolan} and us arises.
First we perform a perturbative expansion of the logarithm in
(\ref{effaction1usual}) over the coupling constant:

\begin{equation}
\Gamma^{\prime}_{\mathit{eff}}[\phi ]=i \int \frac{d^4k}{(2 \pi)^4} \mathrm{tr}
\langle k| \  g \ \phi (\hat{x}) \ S(\hat{p}) + \frac{1}{2} \  g\
\phi (\hat{x}) \ S(\hat{p}) \  g \ \phi (\hat{x}) \ S(\hat{p}) +
... |k \rangle.
\label{perturbation}
\end{equation}

If we wish to reproduce the result of \cite{Dolan}, we do the
derivative expansion of $\phi(\hat{x})$, truncate it to zeroth order and substitute the
constant term $\phi$ in (\ref{perturbation}). Each term of the
expansion depends only on the momentum operator and is diagonal in
momentum space. Therefore the effective action can be resummed as
follows

\begin{eqnarray}
\Gamma^{\prime}_{\mathit{eff}}[ \phi ] &=& i \left\{ g \ \phi \ \int \frac{d^4k}{(2 \pi)^4} 
\ \mathrm{tr} \  \frac{1}{\not\! k - m} + \frac{1}{2}(g\  \phi)^2 \ \int
\frac{d^4k}{(2 \pi)^4} \ \mathrm{tr} \ \left[ \frac{1}{\not\! k - m}
\right]^2 + ... \right\} \int d^4 x \nonumber \\ &=& - i \ \mathrm{tr} \int \frac{d^4k}{(2 \pi)^4} \ln{[1-g \ \phi \
S(k)]} \int d^4 x.
\label{Dolaneffaction}
\end{eqnarray}

\noindent This is the effective action of \cite{Dolan} which, after
some differentiation trick explained therein, yields the effective 
potential (\ref{exact effective potential}). 

However, if we want to find contributions to the effective
action of the form (\ref{effaction2}) with the self-energy given by
(\ref{usual}), then, before performing the derivative expansion, we
should introduce complete sets of intermediate states in
(\ref{perturbation}) and let the momentum and space operators act on
them. This yields

\begin{eqnarray}
\Gamma^{\prime}_{\mathit{eff}}[ \phi ]& =&i \left\{ g \int d^4x \int \frac{d^4k}{(2 \pi )^4}
\  \mathrm{tr} \ \langle k| \phi( \hat{x} ) |x \rangle \ \langle x| S(
\hat{p} ) | k \rangle \right. \nonumber \\
 &+& \left.  \frac{1}{2} g^2 \int d^4x \int d^4y \int \frac{d^4 q}{(2
 \pi )^4 } \int \frac{d^4k}{(2\pi)^4} \ \mathrm{tr} \ \langle k |
 \phi( \hat{x} ) | x \rangle \  \langle x | S( \hat{p} )
 |q \rangle \ \langle q| \phi( \hat{x} ) |y \rangle \ \langle y| S(
\hat{p} ) | k \rangle  + ... \right\} \nonumber \\
& = &i \left\{ g \int d^4x \int \frac{d^4k}{ (2 \pi )^4 }
\ \mathrm{tr} \ \phi(x) \ \frac{1}{\not\! k - m}  \right. \nonumber \\
&+& \left. \frac{1}{2} g^2 \int \frac{d^4 q}{(2
\pi)^4 } \  \tilde{ \phi } (-q) \ \int \frac{d^4k}{(2 \pi)^4}
\ \mathrm{tr} \ \frac{1}{[ \not\! k - m ][ \not\! k + \not\! q - m]} \
\tilde{\phi}(q) + ... \right\}
\end{eqnarray}   

\noindent where $\tilde{\phi}(q)$ is the Fourier transformation of
$\phi(x)$. The final step of our calculation is to perform the
derivative expansion of $\phi(x)$ and keep only the constant term.
Thus we obtain,

\begin{equation}
\Gamma^{\prime}_{\mathit{eff}}[ \phi ] = i \left\{ g \ \phi \ \int \frac{d^4 k}{(2\pi)^4}
 \ \mathrm{tr} \ \frac{1}{ \not\! k - m } \ \int d^4 x + \frac{1}{2} \
 (g \ \phi )^2 \  \lim_{q \rightarrow 0}  \ \int \frac{d^4k}{(2 \pi)^4}
 \ \mathrm{tr} \ \frac{1}{[ \not\! k - m ][ \not\! k + \not\! q - m]}
\ \int d^4x + ... \right\}. 
\label{oureffaction}    
\end{equation}

\noindent As we have already mentioned, at finite temperature, doing
the integration (the Matsubara sum) first and then taking the limit in the second term of (\ref{oureffaction}) is not equivalent to taking the limit of the
integrand first and then performing the integration (the Matsubara sum).
Comparing (\ref{Dolaneffaction}) to (\ref{oureffaction}) we can
clearly see that this difference is due to the interchange of the order with
which we perform the derivative expansion and the action of the
momentum and space operators on the bras and kets. Replacing
$\phi(\hat{x})$ with the constant $\phi$ in the beginning of the
calculation, as Dolan and Jackiw do, means that we lose the operator
behaviour of the $\phi $ field. 

Before proceeding, we shall add a further comment. If we had applied the limit before performing the angular integration in (\ref{usual thermal bubble 1}), we would have found

\begin{eqnarray}
(V_{\mathit{eff}}^{(2)})^{'}&=&-\frac{g^2}{2 \pi^2} \ \int_{|m|}^{\infty} d\w \left\{ \tanh{\frac{\beta \w}{2}} \ \sqrt{\w ^2 -m^2} \left[1- \frac{m^2}{\w ^2} \right] \right. \nonumber \\
 & & \left. + \frac{m^2 \beta}{2 \w} \ \cosh^{-2}{\frac{\beta \w}{2}}
\left[\sqrt{\w^2 - m^2} - \frac{\w a}{2} \ \ln{\left| \frac{\w a +
\sqrt{ \w ^2 - m^2}}{\w a - \sqrt{\w ^2 - m^2}} \right|} \right] \right\} \phi^2 
\end{eqnarray}

\noindent which depends on $a$ and differs from the result of Dolan and Jackiw, as (\ref{usual effective potential}) does. This new result would have an additional property compared to (\ref{usual effective potential}), it would reduce to the result of Dolan and Jackiw in the $\left\{ q_0 \rightarrow 0, \bq \rightarrow 0 \right\}$ limit or equivalently in the $a \rightarrow 0$ limit. This property is frequently mentioned in the literature, see for example \cite{Gribosky}. However it is an artefact of interchanging the limit with the integration and this is why we chose to perform the angular integration first.

\subsection{Comparison with Gribosky and Holstein} 

There is another work \cite{Gribosky} where Gribosky and Holstein 
claim that they have a model which doesn't display the usual
non-analyticity at the origin of the momentum-frequency
space. Actually they employ two ways of proving this, a
non-perturbative way and a perturbative one. However Weldon in his
paper \cite{WeldonMishaps} proves that at least the perturbative way
of Gribosky and Holstein is wrong. The reason is that Feynman
parameterization at finite temperature is not as straightforward as it
is at zero temperature and performing it naively, as Gribosky and
Holstein did, gives misleading results. There is one issue
which still remains open though; where is the mistake in the non-perturbative
 approach of Gribosky and Holstein? Let us examine this matter in some
more detail. The model which the two authors consider is the
following:

\begin{equation}
L[B, \phi ] = L_0 [B(x)] - \frac{1}{2} \ \phi (x) \left( \Box + m^2 + \lambda
 B(x) \right ) \phi (x)       
\end{equation}

\noindent where $L_0[B]$ is the free Klein-Gordon Lagrangian. The
effective Lagrangian and therefore the effective action can be found
from the coincident limit of the Green's function for $\phi(x)$, see
\cite{Schwinger}. Next they write down the differential equation which defines the Green's function:   

\begin{equation}
\left( \Box + m^2 + \lambda B(x) \right ) G(x,x^{'}) = - \delta ( x -
 x^{'} ). \label{difeq}
\end{equation}

\noindent The above equation cannot be solved for a general $B(x)$, it is possible
though to solve it, if we perform the derivative expansion of $B(x)$
around $x^{'}$ and keep terms of up to second order in derivatives.
This method first appeared in paper \cite{DittrichSource} by Dittrich.
Our criticism is that, although at zero temperature there is no problem with this
approximation, at finite temperature it amounts to replacing a 
periodic function with one that is not periodic. In other words $B(x)$
obeys $ B( \tau + \beta ) = B( \tau ) $ but the expansion which is
truncated at second or even at first order in derivatives doesn't and
consequently this non-perturbative method cannot
be applied at finite temperature. The fact that this is the reason for
 obtaining a result which doesn't display the usual problem
of non-uniqueness can be seen from the same paper of
Gribosky and Holstein \cite{Gribosky}! In their perturbative approach they remark
under their equation (3.20b) that `` Thus, we see that extending
$\Pi (p)$ to continuous $p_0$ without ever using $ p_0=2 \pi i
 l / \beta  $ eliminates the non-analytic behaviour we encountered in
$\Pi_R(p)$ '' where $\Pi_R(p)$ is the self-energy which
they found after using (\ref{crucial}) which follows from $ p_0=2 \pi i l/
\beta $ (their $p_0$ is our $q_0$). However, it is
well-known that this expression for $p_0$ is a direct consequence of
the periodicity of $B(x)$ which is lost in the beginning of their
non-perturbative calculation. 

\subsection{Comparison with Evans}

The effects of retaining $\exp{(\beta q_0)}$ after the analytic
continuation where recently re-investigated in papers
\cite{EvansDerivative,EvansUnique} by Evans. Again the
conclusion is that keeping this exponential cures the non-uniqueness
of the thermal effective potential. We can also see this in our
own calculation. As we have already shown in section (3),
 the theory examined there has the
usual problems of non-uniqueness due to the $\Pi^{''}$ piece of the
self-energy. Let us see what happens to $\Pi^{''}$, if we follow
Evans.  Letting $q_0$ be real we recast (\ref{usual thermal bubble 1}) as follows 
\begin{eqnarray}
&& \Pi^{''}= -2 m^2 \ \int \frac{d^3 {\bf k}}{(2 \pi)^3} \ \frac{1}{\W \w }
\ \frac{1}{e^{\beta \w} + 1} \times \nonumber \\
&& \left\{ \frac{ e^{\beta (\W +\w - q_0)} - 1}{ \W + \w - q_0
} \ \frac{1}{ e^{ \beta ( \W - q_0)} + 1} + \frac{ e^{\beta (\W +\w +
q_0)} - 1}{ \W + \w + q_0} \ \frac{1}{ e^{ \beta ( \W + q_0)} + 1} \right.
\nonumber \\
&& \left. \frac{1- e^{\beta (\W -\w + q_0)}}{\W -\w + q_0} \
\frac{e^{\beta \w }}{ e^{ \beta ( \W + q_0)} + 1 } +   \frac{ 1 -
e^{\beta (\W -\w - q_0)}}{\W -\w - q_0} \
\frac{e^{\beta \w }}{ e^{ \beta ( \W - q_0)} + 1 } \right\}.
\label{usual thermal bubble unique}
\end{eqnarray}
\noindent We observe that, because we have retained $q_0$ in the
exponentials appearing in the numerators of the four terms, the
integrand of (\ref{usual thermal bubble unique}) does not have poles
at $ q_0 = \W + \w , - \W - \w , -\W + \w , \W - \w $ anymore. These
poles of the integrand were responsible for the branch points of
(\ref{usual thermal bubble 2}) at $0$ and $4m$ in the complex $s=q_0^2
- \bmq ^2 $ plane as explained in \cite{WeldonRules} and more
generally in \cite{Eden}. We know, for example see 
\cite{WeldonMishaps}, that the branch point at $0$ is associated with the
non-uniqueness of the effective potential and the elimination of this branch point
leads to a unique effective potential. 

However, it seems to us, that this general way out of the problem is
incorrect. The argument which Evans gives for keeping $\exp{\beta
q_0}$ after the analytic continuation is ``The key idea is that such
(derivative) expansions describe field configurations which vary
{\em slowly} in time and space and hence are not thermal equilibrium
configurations (i.e. are not periodic in time). By explicit
calculation, I will show that the retarded thermal Green functions
used in previous analyses are {\em not} relevant to this problem.
These Green functions describe how the system responds to a {\em
sudden} impulse as linear response theory shows.'' ( third paragraph
of \cite{EvansDerivative}). It is not clear to
us what the periodicity of a field has to do with whether it is
slowly or rapidly varying in time. Our point of view is that the
truncated derivative expansion to 1st or higher order in derivatives
 may not be periodic itself but, provided that the periodicity of the
exact field configuration is not forgotten, it provides a good
approximation, when we deal with slowly varying, periodic in time, fields.

Moreover, in all usual physical situations, the Landau terms are expected to cancel each other at the zero temperature limit. If $\exp{\beta q_0} \neq 1$, (\ref{usual thermal bubble 1}) does not have this property unless $|q_0|< \W $. This reassures us that $\exp{\beta q_0}$ must be set equal to $1$.

\subsection{Comparison with Weldon}

It occurred to us that, although we are not convinced by the arguments presented in \cite{EvansDerivative,EvansUnique}, there may be physical situations where the calculation of Evans is correct for a different reason, namely the exact configuration itself for the external field may not be necessarily periodic. In \cite{WeldonRules}, Weldon claims that, although the fields in the loop of the self-energy have to be in thermal equilibrium, it is not necessary to assume that the external field $\phi$ is in thermal equilibrium. Initially, it should be taken to follow a non-equilibrium thermal distribution $f_{0}(q_0)$. At any later time this distribution will be denoted as $f(q_0,t)$. Changes in $f$ result both from the processes $\phi \rightarrow \psi \bar{\psi}$, $\phi \bar{\psi} \rightarrow \bar{\psi}$, $\phi \psi \rightarrow \psi$ which decrease the number of $\phi$'s with rate $f \Gamma_d$ and from the processes $ \psi \bar{\psi} \rightarrow \phi$, $\psi \rightarrow \psi \phi$, $ \bar{\psi} \rightarrow \bar{\psi} \phi$ which increase the number of $\phi$'s with rate $(1+f)\Gamma_i$. Thus $f(q_0,t)$ satisfies:

\begin{equation}
\frac{\partial f}{\partial t}= -f \Gamma_d + (1+f) \Gamma_i.
\end{equation}

\noindent For small departures from equilibrium one finds the solution 

\begin{equation}
f(q_0,t)= \frac{1}{e^{\beta q_0} -1} + c(q_0) \ e^{- \Gamma (q_0) \ t }
\end{equation}

\noindent where $c(q_0)$ is some arbitrary function. Regardless of the distribution specified initially, $f(q_0,t)$ inevitably approaches the equilibrium as $t \rightarrow \infty $. The rate of approach to equilibrium $\Gamma(q_0)$ is related to the imaginary part of the self-energy $\mathrm{Im} \Pi $ through the relation $ \mathrm{Im} \Pi = -q_0 \Gamma (q_0) $. 

Does this mean that the external field is non-periodic and therefore $\exp{\beta q_0} $ should be retained after the analytic continuation? At this point one realizes that the calculation of \cite{WeldonRules} was not performed consistently, since $\exp{\beta q_0}$ was set to $1$. This casts doubts concerning the conclusions of this paper. Furthermore, in our calculation, it is easy to see that retaining the $\exp{\beta q_0} $ leads to vanishing imaginary part of the self-energy.  

The imaginary part of (\ref{usual thermal bubble unique}) is 

\begin{eqnarray}
&& \mathrm{Im} \Pi^{''}_{R}=- \mathrm{Im} \Pi^{''}_{A} = \frac{m^2}{2 \pi q} \int_{|m|}^{\infty} \frac{d\w }{e^{\beta \w} + 1} \ \int_{\W _{-}}^{\W _{+}} d\W \times \nonumber \\
&& \left\{  \left[1- e^{\beta (\W + \w - q_0)}\right] \frac{1}{1+ e^{\beta (\W - q_0)}} \ \delta(\W + \w -q_0) - \left[1- e^{\beta (\W + \w + q_0)}\right] \frac{1}{1+ e^{\beta (\W + q_0)}} \ \delta( \W + \w + q_0) \right. \nonumber \\
&& \left.  + \left[1- e^{\beta (\W - \w + q_0)}\right] \frac{e^{\beta \w}}{1+ e^{\beta (\W + q_0)}} \ \delta(\W - \w + q_0) - \left [1- e^{\beta (\W - \w - q_0)} \right] \frac{e^{\beta \w}}{1+ e^{\beta (\W - q_0)}} \ \delta(\W - \w - q_0) \right\} \nonumber \\
&& =0.
\end{eqnarray}

\noindent Similarly we can show that, if the $\exp{\beta q_0}$ is retained, the imaginary part of (\ref{thermal bubble before}) vanishes.

Consequently, if we apply the conclusions of \cite{WeldonRules} in our case, we reach a contradiction; The imaginary part of the self-energy gives the rate at which the thermal distribution of $\phi$ approaches the equilibrium as $t \rightarrow \infty $. This implies that at any finite time $\phi$ is not in thermal equilibrium and therefore not periodic. However this means that the imaginary part of the self-energy vanishes, there is no approach to equilibrium and the physical interpretation given in \cite{WeldonRules} seems to be meaningless.

\section{Conclusions}

We have shown that in a model, where fermions couple to a
pseudo-scalar field, the effective potential for the pseudo-scalar field can be
found uniquely at finite temperature. We have also shown that this is
not true when the fermion couples to a scalar field, the reason for
that being the non-analytic behaviour which appears in the thermal
bubble diagram. 

We have pointed out that truncating in the beginning of the
calculation the derivative expansion either at the constant term
\cite{Dolan} or at higher order \cite{Gribosky, EvansDerivative,
EvansUnique} gives misleading results. In the former case, because the
operator nature of the background field is lost, and in the latter
case, because the periodicity of the background field is not taken
into account. 

Finally we note that the models we dealt with in sections 2 and 3 can be
considered together to study chiral symmetry restoration at finite
temperature for example in the Lurie model \cite{Wang}, the linear
$\sigma$ model \cite{Gell} in its broken
chiral symmetry phase and in the Nambu-Jona-Lasinio model
\cite{NJL} expressed in terms of auxiliary fields. In general,
whenever finite temperature symmetry restoration is discussed by
employing non-perturbative results for the effective potential,
 these may not match the perturbative results.
 Therefore the question of symmetry restoration at finite
temperature should be reanalyzed keeping in mind the non-analyticity
of some graphs. Work on this and other related issues is in progress.

\section*{Acknowledgements}

The author wishes to thank Prof. I.J.R. Aitchison and Dr. M.B. Hott for
 valuable discussions. This work was partially supported by PPARC (UK)
  under studentship no. PPA/S/S/1997/02518.

%\bibliography{vjcw}

\newpage

\begin{thebibliography}{10}

\bibitem{WeldonMishaps}
H.~A. Weldon, Phys. Rev. {\bf D47},  594  (1993).

\bibitem{Das1}
A. Das, {\em Finite Temperature Field Theory} (World Scientific, Singapore,
  1997).

\bibitem{Kajantie}
K. Kajantie and J. Kapusta, Ann. Phys. {\bf 160},  477  (1985).

\bibitem{WeldonCovariant}
H.~A. Weldon, Phys. Rev. {\bf D26},  1394  (1982).

\bibitem{Jackson}
J.~D. Jackson, {\em Classical Electrodynamics} (Wiley, New York, 1975).

\bibitem{Zuk}
I.~J.~R. Aitchison and J.~A. Zuk, Ann. Phys. {\bf 242},  77  (1995).

\bibitem{Hott}
A. Das and M. Hott, Phys. Rev. {\bf D50},  6655  (1994).

\bibitem{Tsuneto}
E. Abrahams and T. Tsuneto, Phys.Rev. {\bf 152},  416  (1966).

\bibitem{Kalashnikov}
O.~K. Kalashnikov and V.~V. Klimov, Sov. J. Nucl. Phys. {\bf 31},  699  (1980).

\bibitem{Klimov}
V.~V. Klimov, Sov. J. Nucl. Phys. {\bf 33},  934  (1981).

\bibitem{WeldonQuark}
H.~A. Weldon, Phys. Rev. {\bf D},  2789  (1982).

\bibitem{Braaten}
E. Braaten and R.~D. Pisarski, Nucl. Phys. {\bf B337},  569  (1990).

\bibitem{FrenkelHigh}
J. Frenkel and J.~C. Taylor, Nucl. Phys. {\bf B334},  199  (1990).

\bibitem{RebhanAnalytical}
A. Rebhan, Nucl. Phys. {\bf B368},  479  (1992).

\bibitem{RebhanCollective}
A. Rebhan, Nucl. Phys. {\bf B351},  706  (1991).

\bibitem{FrenkelHard}
J. Frenkel and J.~C. Taylor, Z. Phys. {\bf C49},  515  (1991).

\bibitem{Fujimoto}
Y. Fujimoto and H. Yamada, Z. Phys. {\bf C37},  265  (1988).

\bibitem{EvansZero}
T.~S. Evans, Z. Phys. {\bf C36},  153  (1987).

\bibitem{WeldonRules}
H.~A. Weldon, Phys. Rev. {\bf D28},  2007  (1983).

\bibitem{Vokos}
P. Arnold, S. Vokos, P. Bedaque, and A. Das, Phys. Rev. {\bf D47},  4698
  (1993).

\bibitem{MetikasHott1}
M. Hott and G. Metikas, hep-ph/9812227  .

\bibitem{Dolan}
L. Dolan and R. Jackiw, Phys. Rev. {\bf D9},  3320  (1974).

\bibitem{Gribosky}
P.~S. Gribosky and B.~R. Holstein, Z. Phys. {\bf C47},  205  (1990).

\bibitem{Schwinger}
J. Schwinger, Phys.Rev. {\bf 82},  664  (1951).

\bibitem{DittrichSource}
W. Dittrich, Fortsch. Phys. {\bf 26},  289  (1978).

\bibitem{EvansDerivative}
T.~S. Evans, hep-ph/9808382  .

\bibitem{EvansUnique}
T.~S. Evans, hep-ph/9808383  .

\bibitem{Eden}
Eden, Landshoff, Olive, and Polkinghorne, {\em The Analytic S-Matrix}
  (Cambridge University Press, Cambridge, 1966).

\bibitem{Wang}
E. ke~Wang and J. rong Li, Phys. Rev. {\bf D44},  3680  (1991).

\bibitem{Gell}
M. Gell-Mann and M. Levy, Nuovo Cim. {\bf 16},  705  (1960).

\bibitem{NJL}
Y. Nambu and G. Jona-Lasinio, Phys. Rev. {\bf 122},  345  (1961).

\end{thebibliography}

\end{document}